\documentclass[submission,copyright,creativecommons]{eptcs}

\providecommand{\event}{LFMTP 2018} 
\usepackage{underscore}           
\usepackage{microtype}
\usepackage{amssymb}
\usepackage{amsmath}
\usepackage{amsthm}
\usepackage{bbm}
\usepackage{textgreek}
\usepackage{epigraph}
\usepackage{fancyhdr}
\usepackage{fancyvrb}
\usepackage{verbatim}
\usepackage[utf8]{inputenc}
\usepackage{newunicodechar}
\usepackage{autofe}
\usepackage{tipa}
\usepackage{tikz}
\usepackage{extpfeil}

\newcommand{\rightsquigarrowdbl}{\rightsquigarrow\mathrel{\mkern-9mu}\shortrightarrow}

\newcommand{\upd}{\ensuremath{\prec\hspace{-.4em}+}}
\newcommand{\aum}{\upd}
\newcommand{\hap}{\ensuremath{\twoheadrightarrow_{hap}}}
\newcommand{\shap}{\ensuremath{\longrightarrow_{hap}}}
\newcommand{\bet}{\ensuremath{\twoheadrightarrow_{\beta}}}
\newcommand{\sbet}{\ensuremath{\longrightarrow_{\beta}}}
\newcommand{\leftmost}{\ensuremath{\twoheadrightarrow_{l}}}
\newcommand{\sleftmost}{\ensuremath{\longrightarrow_{l}}}

\newcommand{\sta}{\ensuremath{\twoheadrightarrow_{st}}}

\newcommand{\sst}{\ensuremath{\rightarrow_{st}}}
\newcommand{\sust}{\ensuremath{\bullet\ }}
\newcommand{\alfa}{\ensuremath{\sim_{\alpha}}}
\newcommand{\lambdabar}{\mbox{\textipa{\textcrlambda}}}
\DeclareUnicodeCharacter{8988}{\ensuremath{\ulcorner}}
\DeclareUnicodeCharacter{8989}{\ensuremath{\urcorner}}
\DeclareUnicodeCharacter{8803}{\ensuremath{\overline{\equiv}}}
\DeclareUnicodeCharacter{411}{\textipa{\textcrlambda}}
\DeclareUnicodeCharacter{65288}{(}
\DeclareUnicodeCharacter{65289}{)}
\DeclareUnicodeCharacter{8615}{\ensuremath{\downarrow}}
\DeclareUnicodeCharacter{8788}{\ensuremath{\coloneqq}}
\DeclareUnicodeCharacter{8336}{\ensuremath{_a}}
\DeclareUnicodeCharacter{8799}{\ensuremath{\overset{?}{=}}}
\DeclareUnicodeCharacter{8759}{\ensuremath{\dblcolon}}
\DeclareUnicodeCharacter{8718}{\ensuremath{\square}}
\DeclareUnicodeCharacter{10239}{\sim\joinrel\leadsto}
\DeclareUnicodeCharacter{2206}{$\textDelta$}
\DeclareUnicodeCharacter{2200}{$\forall$}
\DeclareUnicodeCharacter{21A7}{$\downarrow$}
\DeclareUnicodeCharacter{2115}{$\mathbb{N}$}
\DeclareUnicodeCharacter{2192}{$\rightarrow$}
\DeclareUnicodeCharacter{019B}{$\lambdabar$}
\DeclareUnicodeCharacter{2262}{$\not\equiv$}
\DeclareUnicodeCharacter{2261}{$\equiv$}
\DeclareUnicodeCharacter{2218}{$\circ$}
\DeclareUnicodeCharacter{227A}{$\prec$}
\DeclareUnicodeCharacter{225F}{$\stackrel{?}{=}$}
\DeclareUnicodeCharacter{00D7}{$\times$}
\DeclareUnicodeCharacter{27FF}{$\rightsquigarrow$}
\DeclareUnicodeCharacter{21C2}{$\downharpoonright$}
\DeclareUnicodeCharacter{2219}{$\bullet$}
\DeclareUnicodeCharacter{223C}{$\sim$}
\DeclareUnicodeCharacter{00AC}{$\neg$}
\DeclareUnicodeCharacter{2080}{$_{0}$}
\DeclareUnicodeCharacter{2228}{$\vee$}
\DeclareUnicodeCharacter{2245}{$\cong$}
\DeclareUnicodeCharacter{27F6}{$\longrightarrow$}
\DeclareUnicodeCharacter{2265}{$\geq$}
\DeclareUnicodeCharacter{2227}{$\wedge$}
\DeclareUnicodeCharacter{2193}{$\downarrow$}
\DeclareUnicodeCharacter{2264}{$\leq$}
\newunicodechar{∀}{$\forall$}
\newunicodechar{↧}{$\downarrow$}
\newunicodechar{ℕ}{$\mathbb{N}$}
\newunicodechar{∼}{$\sim$}
\newunicodechar{ƛ}{$\lambdabar$}
\newunicodechar{⟶}{$\longrightarrow$}
\newunicodechar{≥}{$\geq$}
\newunicodechar{₀}{$_{0}$}
\newunicodechar{∨}{$\vee$}
\newunicodechar{⟿}{$\rightsquigarrow$}
\newunicodechar{⇂}{$\downharpoonright$}
\newunicodechar{∙}{$\bullet$}
\newunicodechar{≡}{$\equiv$}
\newunicodechar{≺}{$\prec$}
\newunicodechar{≅}{$\cong$}
\newunicodechar{∧}{$\wedge$}
\newunicodechar{≤}{$\leq$}
\newunicodechar{≢}{$\not\equiv$}
\newunicodechar{∘}{$\circ$}
\newunicodechar{≟}{$\stackrel{?}{=}$}
\newunicodechar{Λ}{$\Lambda$}
\newunicodechar{ƛ}{$\lambdabar$}
\newunicodechar{Σ}{$\Sigma$}
\newunicodechar{ι}{$\iota$}
\newunicodechar{σ}{$\sigma$}
\newunicodechar{χ}{$\chi$}
\newunicodechar{α}{$\alpha$}
\newunicodechar{β}{$\beta$}
\newunicodechar{λ}{$\lambda$}
\DefineVerbatimEnvironment{code}{Verbatim}{fontsize=\small}

\title{Formalization in Constructive Type Theory of the Standardization Theorem for the Lambda Calculus using Multiple Substitution}

\author{Martín Copes \qquad\qquad Nora Szasz \qquad\qquad Álvaro Tasistro
\institute{Universidad ORT Uruguay}
\email{\texttt{\{copes,szasz,tasistro\}@ort.edu.uy}}
}

\def\titlerunning{Formalization in Constructive Type Theory of the Standardization Theorem} 
\def\authorrunning{M. Copes, N. Szasz and A. Tasistro}

\begin{document}

\maketitle

\begin{abstract}
We present a full formalization in Martin-L\"of's Constructive Type Theory of the Standardization Theorem for the Lambda Calculus using first-order syntax with one sort of names for both free and bound variables and Stoughton's multiple substitution.
Our formalization is based on a proof by Ryo Kashima, in which a notion of $\beta$-reducibility with a standard sequence is captured by an inductive relation.
The proof uses only structural induction over the syntax and the relations defined, which is possible due to the specific formulation of substitution that we employ.
The whole development has been machine-checked using the system Agda.
\end{abstract}

\section{Introduction} 
\label{intro} 

In~\cite{COPELLO2016} a formalization of the Lambda Calculus in Martin-L\"of's Constructive Type Theory is presented, which uses first-order syntax with one sort of names for both free and bound variables that does not identify $\alpha$-convertible terms, and a multiple substitution operation introduced  by Stoughton in~\cite{stoughton}. The approach enables the authors to prove in a completely formal and  quite elegant way significant results about the metatheory of the Lambda Calculus, namely the Church-Rosser Theorem and Subject Reduction for the simply typed Lambda Calculus à la Curry. The authors developed a library~\cite{ernius:lib} with definitions and lemmas for implementing and manipulating substitutions that was key for achieving the mentioned results, in particular by using only simple standard methods of structural induction on terms and reduction relations.

In the present work we extend the above mentioned metatheoretical study by proving the Standardization Theorem for $\beta$-reduction, which we further use to prove that the leftmost-outermost reduction strategy always finds the normal form of a term provided that it exists. 
The Standardization Theorem is a well-known result in the Lambda Calculus that was first proved by Curry and Feys in~\cite{curry}. 
It states that if a term $M$ $\beta$-reduces to a term $N$, then there exists a standard $\beta$-reduction sequence from $M$ to $N$. A reduction sequence is considered standard if successive redexes are contracted from left to right (regarding the linear syntax) possibly with some jumps. 

The proof hereby formalized is the one given by Ryo Kashima in~\cite{kashima} where the notion of $\beta$-reducibility with a standard sequence is captured by an inductive relation in very much the same way as minimal complete developments are captured by the so-called \emph{parallel} reduction relation in e.g. Tait and Martin-L\"of's method for proving the Church-Rosser theorem.
This allows for an elegant inductive development  
as opposed to basing the proofs on notions like residuals and finite developments as in  the classical proofs by Curry-Feys and Barendregt \cite{curry, bar:84}. 



All the definitions and proofs that appear in this article have been 
machine-checked with the system Agda~\cite{agda}. 
In the subsequent text we will mix Agda code and (informal) proofs in English with a considerable level of detail so that they serve for clarifying their formalization.  The complete code is available at {\tt https://github.com/mcopes73/standardization-agda}.  

In section~\ref{prelim}  we present the basic concepts of the Lambda Calculus, together with some definitions and results from the library produced in~\cite{COPELLO2016} on which our work is based, as well as extensions thereof. 
In section~\ref{proof} we present the proof of the Standardization Theorem.
In section~\ref{leftmost} we present the proof of the Leftmost Reduction Theorem for $\beta$-reduction. 
In section~\ref{concl} we compare  our development with other similar efforts in the literature, and present our overall conclusions.

\section{Preliminaries}
\label{prelim}In what follows we will introduce the main definitions and results in~\cite{COPELLO2016,ernius:lib} that are previous to this work and are used in our formalization.
We present the definitions directly using Agda code along with informal explanations, while the proofs are written in English to ease their reading.
A certain degree of familiarity with the Agda syntax or at least with that of functional languages like Haskell is assumed. 

We shall start by defining $\lambda$-terms using the same set of names for both bound and free variables. We use natural numbers to name variables for sake of concreteness.

\begin{code}
V = ℕ
data Λ : Set where
  v    : V → Λ 
  _·_  : Λ → Λ → Λ 
  ƛ    : V → Λ → Λ
\end{code}

Agda is pretty liberal with regard to the naming of functions and the positions of their arguments. Notice the notation for declaring the infix application constructor, i.e. \ {\tt \_·\_}. This underscore notation is extended to mixfix operators.

The classical notions of {\em free} and {\em fresh} (not free) variable in a term, which are denoted by $*$ \ and $\#$ \ respectively, are defined as binary relations between variables and terms in the usual way (we omit the definitions for reasons of space):
\begin{code}
data _*_ : V → Λ → Set
data _#_ : V → Λ → Set 
\end{code}
%

{\em Substitutions} are {\em identity-almost-everywhere} functions associating a term to every variable. We can generate every concrete substitution by starting up from the empty substitution $\iota$ \ that maps each variable to itself as a term, and employing the update operation \upd, such that if $\sigma$ is a substitution, then $\sigma \upd (x,M)$ is the substitution equal to $\sigma$ \ everywhere except at $x$, where it yields $M$:

\begin{code}
Σ = V → Λ

ι : Σ 
ι = id ∘ v

_≺+_ : Σ → V × Λ → Σ
(σ ≺+ (x , M)) y with x ≟ y
... | yes _ = M
... | no  _ = σ y
\end{code}

Notice that in the definition of {\tt ≺+ } we use the {\tt with} construct, which allows us to perform pattern matching on the result of evaluating the expression {\tt x ≟ y}.
This expression decides the equality between the variables {\tt x} and {\tt y} and has type {\tt Dec $\equiv$}, whose constructors are {\tt yes} and {\tt no} applied to the corresponding proof objects. 

In general, we shall consider properties concerning the substitutions for the free variables of a term $M$, i.e. their {\em restrictions} to such variables. The type of restrictions {\tt R} is defined as: \ {\tt R = Σ × Λ}, and we note in the informal language such a restriction as $\sigma\downharpoonright M$. This means that we are restricting the substitution $\sigma$ to the free variables of $M$ only.
We will also use the following notion: $x\ \#\ (\sigma\downharpoonright M)$, which stands for {\em $x$ fresh in  the  $\sigma$-value of every free variable of $M$}:

\begin{code}
_#⇂_ : V → R → Set
x #⇂ (σ , M) = (y : V) → y * M → x # (σ y)
\end{code}

The application of substitution $\sigma$ to the term $M$ is noted  $M\, \bullet\, \sigma$, and it is defined by \emph{structural recursion} on $M$. 
The fact that structural recursion is sufficient for stating this very concrete definition is a (very welcome) non-trivial consequence of the employment of multiple substitutions.

\begin{code}
_∙_ : Λ → Σ → Λ
(v x) ∙ σ = σ x
(M · N) ∙ σ = (M ∙ σ) · (N ∙ σ)
(ƛ x M) ∙ σ = ƛ y (M ∙ (σ ≺+ (x , v y)))
     where y = χ (σ , ƛ x M)
\end{code}

Notice the last line of the definition: when performing a substitution over an abstraction, the bound variable $x$ is always replaced with a new one. This new variable $y$ is obtained by means of a choice function $\chi$, such that $\chi(\sigma,M)\ \#\ (\sigma\downharpoonright M)$. 
In this way, $y$ does not capture any of the names introduced into its scope by effect of the substitution\footnote{In fact, $\chi$ is implemented by just finding the first variable not free in the given restriction.}.
When reasoning with substitutions, this uniform renaming of bound variables allows us to avoid case analyses; it also has other nice consequences, to be noticed shortly.
For the sake of readability, we define the single substitution of a term $N$ for a variable $x$ in $M$ with the traditional notation $M[x:=N]$.
\begin{code}
_[_:=_] : Λ -> V -> Λ -> Λ
M [ x := N ] = M ∙ (ι ≺+ (x , N))
\end{code}

Alpha-conversion (\alfa) is defined as the following inductive binary relation on terms:

\begin{code}
data _∼α_ : Λ → Λ → Set where
  ∼v : {x : V} → (v x) ∼α (v x)
  ∼· : {M M' N N' : Λ} → M ∼α M' → N ∼α N' → M · N ∼α M' · N'
  ∼ƛ : {M M' : Λ}{x x' y : V} → y # ƛ x M → y # ƛ x' M' 
                               → M [x := v y] ∼α M' [x' := v y]
                               → ƛ x M ∼α ƛ x' M'
\end{code}
Arguments to a function declared between braces {\tt \{\,\}} are optional and in subsequent applications of the function in question they  are inferred by the type-checker. 
The first two constructors above implement the classical rules for variables and application. The last constructor states that two abstractions are $\alpha$-convertible if and only if their bodies are $\alpha$-convertible after replacing the bound variables with a common fresh name. From this definition it follows that \alfa\ is an equivalence relation, as shown in~\cite{COPELLO2016}. 
As it is the case in~\cite{stoughton}, $\alpha$-equivalent terms become identical when submitted to the same substitution. This is due to the fact that abstractions are uniformly renamed, and that the new name chosen by the $\chi$ function is determined only by the restriction of the substitution to the free variables of the terms, which is the same one if the terms are $\alpha$-equivalent. 
This is proven in~\cite{COPELLO2016}, and we just mention the corresponding lemma here:

\begin{code}
lemmaM∼M'→Mσ≡M'σ : {M M' : Λ}{σ : Σ} → M ∼α M' → M ∙ σ ≡ M' ∙ σ
\end{code}

\noindent From now on we present definitions and results not included in the library~\cite{ernius:lib}. 

Firstly,  we have proven that this definition of alpha equivalence is decidable:

\begin{code}
_∼α?_ : ∀ A B -> Dec (A ∼α B)
\end{code}


Given a binary relation $\rightsquigarrow$, we define its $\alpha$-reflexive-transitive closure as follows:
\begin{code} 
data α-star (⟿ : Rel) : Rel where
    refl : ∀{M} → α-star ⟿ M M
    α-step : ∀{M N N'} → α-star ⟿ M N' → N' ∼α N → α-star ⟿ M N
    append : ∀ {M N K} → α-star ⟿ M K → ⟿ K N → α-star ⟿ M N
\end{code}

\noindent where {\tt Rel} is the type of binary relations over terms. 

\noindent This is the kind of closure that will be applied to our one-step reduction relations. It represents sequences of $\rightsquigarrow$ steps allowing $\alpha$ conversions, which have to be made explicit because we are dealing with concrete terms, i.e. terms not identified under $\alpha$ conversion.
In informal notation we shall write the $\alpha$-reflexive-transitive closure of a relation with the classical two-headed arrow. 
From the definition given we can easily prove that, for any relation $\rightsquigarrow$, $M \rightsquigarrow N$ implies $M \rightsquigarrowdbl N$, and that $\rightsquigarrowdbl$ is transitive. The first proof is straightforward using the constructors {\tt append} and {\tt refl}. Transitivity is proven by induction on the definition of {\tt α-star}.
Therefore we have, in Agda:

\begin{code}
α-star-singl : ∀{⟿ M N} -> ⟿ M N -> α-star ⟿ M N 
α-star-trans : ∀{⟿ M N K} -> α-star ⟿ M K -> α-star ⟿ K N -> α-star ⟿ M N
\end{code}


Following Kashima~\cite{kashima}, we define $\beta$-contraction taking into account the position where the contracted redex appears in the term relative to the other redexes. We start by defining two auxiliary functions: 
 {\tt isAbs} is a predicate that decides whether a term is an abstraction and {\tt countRedexes} a function that counts the number of $\beta$-redexes in a term. 
\begin{code}
data isAbs : Λ -> Set where
  abs : forall {x M} -> isAbs (ƛ x M)
\end{code}

We need to prove that {\tt isAbs} is decidable before being able to define {\tt countRedexes}, since the number of redexes for an application depends on whether the left term is an abstraction.
The proof is straightforward:

\begin{code}
isAbs? : (M : Λ) -> Dec (isAbs M)
\end{code}
Using this property we can define {\tt countRedexes} as follows:

\begin{code}
countRedexes : Λ -> ℕ
countRedexes (v _) = 0
countRedexes (M · N) with isAbs? M
... | yes _ = suc (countRedexes M + countRedexes N)
... | no _ = countRedexes M + countRedexes N
countRedexes (ƛ _ M) = countRedexes M  
\end{code}

Considering the linear syntax of terms, redexes will be numbered in a left-to-right fashion, starting from zero. We shall start by defining the {\em contraction of the $n$-th redex} as a relation between terms depending on the natural number $n$. 

\begin{code}
data _β_@_ : Λ -> Λ -> ℕ -> Set where
  outer-redex : ∀ {x A B} -> ((ƛ x A) · B) β (A [ x := B ]) @ 0
  appNoAbsL : ∀ {n A B C} -> A β B @ n -> ¬ isAbs A -> (A · C) β (B · C) @ n
  appAbsL : ∀ {n A B C} -> A β B @ n -> isAbs A -> (A · C) β (B · C) @ (suc n)
  appNoAbsR : ∀ {n A B C} -> A β B @ n -> ¬ isAbs C
              -> (C · A) β (C · B) @ (n + countRedexes C)
  appAbsR : ∀ {n A B C} -> A β B @ n -> isAbs C
              -> (C · A) β (C · B) @ (suc (n + countRedexes C))
  abs : ∀ {n x A B} -> A β B @ n -> (ƛ x A) β (ƛ x B) @ n
\end{code}

\noindent The {\tt outer-redex} constructor allows the contraction of the outermost redex,  numbered as the one at position zero. The next four constructors are used to perform contractions inside applications. In order to determine the number of the redex contracted we need to identify whether the left hand side term of the application is an abstraction or not (which is necessary to know whether we are stepping over a redex to reduce an inner one). Finally, the {\tt abs} constructor allows contractions inside an abstraction. 




One-step \emph{$\beta$-reduction} (\sbet) \ from $M$ to $N$ can now be defined as the existence of a natural number $n$ such that $N$ can be obtained by contracting the $n$-th redex from $M$.  We use  Agda's dependent ordered pair constructor $\Sigma$ \ to express existential quantification. 

\begin{code}
_⟶β_ : Λ -> Λ -> Set
M ⟶β N = Σ ℕ (\n -> M β N @ n)
\end{code}
It is easily proven by structural induction that this definition  is equivalent to the compatible (with the syntactic constructors) closure of ordinary $\beta$-contraction.


One interesting result that will be useful in our  development is the following $\alpha$-$\beta$ compatibility property:
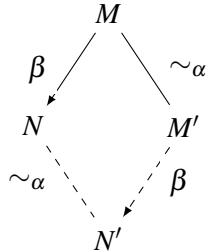
\begin{center}
\begin{tikzpicture}[>=latex]  
\node (M) at (2,2.5) {$M$};
\node (N) at (1,1) {$N$};
\node (MM) at (3,1) {$M'$};
\node (NN) at (2,-0.5) {$N'$};
 \draw [->] (M) --node[left=5pt,fill=white] {$\beta$} (N) ;
 \draw [-] (M) -- node[right=5pt,fill=white] {$\alfa$} (MM) ;
 \draw [dashed,->] (MM) -- node[right=5pt,fill=white] {$\beta$}  (NN) ;
 \draw [dashed,-] (N) --node[left=5pt,fill=white] {$\alfa$} (NN) ;
\end{tikzpicture}
\end{center}
which we state in Agda as the following lemma:
\begin{code}
lem-βα : ∀{M N M'} -> M ⟶β N -> M ∼α M'-> Σ Λ (λ N' -> (M' ⟶β N') ∧ (N' ∼α N))

\end{code}

We finally introduce $\beta$-reduction $\bet$ as the $\alpha$-reflexive-transitive closure of the contraction $\sbet$: 
\begin{code}
_→→β_ : Λ -> Λ -> Set
_→→β_ = α-star (_⟶β_)
\end{code}

\section{The Standardization Theorem}
\label{proof}

In the present section we show the formalization of the Standardization Theorem in Constructive Type Theory that follows the proof given by Kashima in~\cite{kashima}. 
For the sake of clarity, some lemmas are presented in a different order than the one proposed by Kashima. Nonetheless, the formalized results and definitions are the same unless otherwise stated.

\subsection{Standard Reduction Sequences}

A {\em reduction sequence} is a sequence of terms $M_0, M_1,..., M_n$ such that $M_{i+1}$ is obtained from $M_i$ by the contraction of some redex, i.e., $(\forall\ i \in {0...n{-}1}) \ M_{i}\ \sbet M_{i+1}$.
We call a  reduction sequence \emph{standard} if and only if subsequent steps are non decreasing in the number of the redex contracted. 
Kashima defines a standard beta reduction sequence as:
$\textrm{If } M_0\ \xrightarrow{n_1}_\beta  M_1\ \xrightarrow{n_2}_\beta\ ... \ \xrightarrow{n_k}_\beta \ M_k\  \textrm{ then } \ n_1\leq n_2 \leq ... \leq n_k$, 
where $M \xrightarrow{n}_\beta  N$ is the $\beta$-contraction of the $n$-th redex in $M$, which we note $M\,\beta\,N\,@\,n$ in our development.

We implement this notion in Agda by defining a relation indexed on a natural number that keeps track of the lower bound allowed for the next redex to be contracted.
\begin{code}
data seqβ-st (M : Λ) : (N : Λ) -> ℕ -> Set where
  nil : seqβ-st M M 0
  α-step : ∀ {n K N} -> seqβ-st M K n -> K ∼α N -> seqβ-st M N n
  β-step : ∀ {K n n₀ N} -> seqβ-st M K n -> K β N @ n₀ -> n₀ ≥ n -> seqβ-st M N n₀
\end{code}

The relation is reflexive and allows for $\alpha$-steps, which do not appear explicitly in Kashima's definition because the latter relies on the (informal) syntactic identification of $\alpha$ convertible terms.
The ``three dots'' of Kashimas's sequence $M_0, M_1,..., M_k$ are implemented as follows: we can append a term to the reduction sequence provided that it can be obtained by the contraction of a redex at a position greater than or equal to the current lower bound. 
Using this relation the Standardization Theorem can be precisely stated as the existence of a standard sequence between two terms among which there is a $\beta$-reduction:
\begin{code}
standardization : ∀{M N} -> M →→β N -> Σ ℕ (λ n -> seqβ-st M N n)
\end{code}

\subsection{Two Useful Reduction Relations}

The next step is to capture the existence of a standard sequence as an inductively defined reduction relation between terms. To this end, Kashima introduces two auxiliary one-step reduction relations:

\sleftmost\ stands for {\em leftmost reduction} and corresponds to the contraction of the leftmost  redex, i.e. the one at position zero: 
\begin{code}
data _⟶l_ : Λ -> Λ -> Set
    M ⟶l N =  M β N @ 0
\end{code}

$\shap$ stands for {\em head contraction in application} and represents the contraction of the redex in the head position of a chain of applications, i.e.: $(\lambda x M_0) M_1 M_2... M_n\ \shap\ M_0 [x:=M_1]\ M_2...M_n$.
We define this relation in Agda as follows:
\begin{code}
data _⟶hap_ : Λ -> Λ -> Set where
  hap-head : ∀{x A B} -> (ƛ x A) · B ⟶hap (A [ x := B ])
  hap-chain : ∀{C A B} -> A ⟶hap B -> (A · C) ⟶hap (B · C)
\end{code}

Now \leftmost\ and \hap\ are defined as the $\alpha$-reflexive-transitive closures of \sleftmost \ and \shap \ respectively. 

\begin{code}
_→→hap_ : Λ -> Λ -> Set
_→→hap_ = α-star (_⟶hap_)

_→→l_ : Λ -> Λ -> Set
_→→l_ = α-star (_⟶l_)
\end{code}


The first two lemmas state that  head reduction in application \hap\ \ is compatible with application and substitution respectively.

\begin{code}
hap-app-r : ∀{M N P} -> M →→hap N -> M · P →→hap N · P
\end{code}

\begin{proof} By induction on the definition of $M\hap N$. 
\begin{itemize}
\item Case {\tt refl}: We have to prove $(M\ P) \hap (M\ P)$, which follows  by  {\tt refl}.
\item Case {\tt α-step}: Assume that  $M \hap N$ follows from $M \hap N'$ and $N' \alfa N$. Then, we obtain $M\ P \hap N'\ P$ from the induction hypothesis, and since $N'\ P \alfa N\ P$, we construct our goal using {\tt st-alpha}.
\item Case {\tt append}: Assume  $M \hap N$ follows from $M \hap K$ and $K \shap N$. Then we can obtain $M\ P \hap K\ P$ from the induction hypothesis and $K\ P \shap N\ P$ from rule {\tt hap-chain} applied to $K \shap N$. From these, we construct our goal using  {\tt append}.
\end{itemize}
\vspace*{-3.5ex}
\end{proof}

In order to prove that substitution preserves the head reduction relation, we need two lemmas from the substitution library~\cite{ernius:lib}.
The first one  states that substituting $y$ for $x$ and then $N$ for $y$ yields a result $\alpha$-equivalent to substituting $N$ for $x$, provided $y$ is fresh enough. The second one is a form of the substitution composition lemma: 
\begin{code}
corollary1SubstLemma : {x y : V} {σ : Σ}{M N : Λ} → y #⇂ (σ , ƛ x M) 
    → ((M ∙ (σ ≺+ (x , v y))) [y := N]) ∼α  (M ∙ (σ ≺+ (x , N)))

corollary1Prop7 : {M N : Λ}{σ : Σ}{x : V}
    → M ∙ (σ ≺+ (x , N ∙ σ)) ≡ (M [x := N]) ∙ σ
\end{code}


Now we prove that substitution preserves \shap\ up to \alfa:
\begin{code}
lem-hap-subst : ∀{σ M N} -> M ⟶hap N 
    ->  Σ Λ (λ N' -> ((M ∙ σ) ⟶hap N') ∧ (N' ∼α (N ∙ σ)))
\end{code}
\begin{proof}
By induction on the definition of $M \shap N$
\begin{itemize}
\item Case {\tt hap-head}: 
We want to prove that
$((ƛ x\ A)\ B)\ \sust \sigma\ \shap\ N\ \wedge\ N \alfa (A [x:=B])\ \sust \sigma$, for some term N.
Starting from the left hand side:\\
$((ƛ x\ A)\ B)\ \sust \sigma$\\
$\equiv$ (Def. $\sust$)\\
$(ƛ y\ A\ \sust (\sigma \aum(x, y)))\ (B\ \sust \sigma)$ where $y = \chi(\sigma,\ ƛ x\ A)$\\
$\shap$ ({\tt hap-head})\\
$(A\ \sust (\sigma \aum (x, y)))\ [y:= B\ \sust \sigma]$\\
$\alfa$ ({\tt corollary1substLemma}, $y\ \#\ (\sigma,\ ƛ x\ A)$)\\
$A\ \sust (\sigma \aum(x,\ B\ \sust \sigma))$\\
$\equiv$ ({\tt corollary1Prop7})\\
$(A\ [x:= B])\ \sust \sigma$
\item Case {\tt hap-chain}: We need to prove that there exists a term $K$ such that $(M\ P)\ \sust\, \sigma\ \shap K \ \wedge\ K \alfa (N\ P)\, \sust \sigma$, assuming $M \sust \sigma \shap N \sust \sigma$.
This follows directly from rule {\tt hap-chain} applied to $M\ \sust \sigma\ \shap N\ \sust \sigma$ and the definition of \ $\sust$.
\end{itemize}
\vspace*{-3.5ex}
\end{proof}

Kashima originally formulates the previous result just for single substitutions, i.e., of the form $[x:=P]$. 
Our result using multiple substitutions will allow us to rely only on structural induction in our proofs, as we shall see later. 
We can easily extend the previous result to \hap: 
\begin{code}
hap-subst : ∀{M N σ} -> M →→hap N -> (M ∙ σ) →→hap (N ∙ σ)
\end{code}
\begin{proof}
By induction on $M\hap N$:
\begin{itemize}
\item Case {\tt refl}: Direct using {\tt refl}.
\item Case {\tt α-step}: Assume  $M \hap N'$ and $N' \alfa N$. Then, we obtain $M \sust \sigma \hap N'\sust \sigma$ from the induction hypothesis, and by {\tt lemmaM$\sim$M'→Mσ≡M'σ} mentioned in Section~\ref{prelim}, we have that $N'\sust \sigma \equiv N \sust \sigma$, so we construct our goal using the {\tt α-step} rule, since \alfa\ is reflexive.  
\item Case {\tt append}: Assume  $M \hap N$ follows from $M \hap K$ and $K \shap N$. \\Then we can obtain $M\ \sust\ \sigma \hap K\ \sust\ \sigma$ from the induction hypothesis and $(\exists N') (K\ \sust\ \sigma \shap N'\ \wedge\ N' \alfa N\ \sust\ \sigma)$ from the previous lemma ({\tt lem-hap-subst}) applied to $K \shap N$. From this, using {\tt α-star-single} we construct $K\ \sust\ \sigma \hap N'$ and since $N' \alfa N$ we obtain $K\ \sust\ \sigma \hap N$ from rule {\tt $\alpha$-step}. Finally, we prove our goal from the transitivity of $\hap$.
\end{itemize}
\vspace*{-3.5ex}
\end{proof}

Finally, notice that head reduction in application implies leftmost reduction:
\begin{code}
lem-hap→l : ∀ {M N} -> M ⟶hap N -> M ⟶l N
\end{code}

\noindent And therefore the same inclusion holds for their $\alpha$-reflexive-transitive closures:

\begin{code}
hap→l : ∀{M N} -> M →→hap N -> M →→l N
\end{code}

\subsection{Standard Reduction}

Using \hap, Kashima characterizes the existence of a standard sequence as a further reduction relation \sta, which stands for {\em standard reduction}, as follows:
\begin{code}
data _→→st_ (L : Λ) : Λ -> Set where
  st-var : ∀{x} -> L →→hap (v x) -> L →→st (v x)
  st-app : ∀{A B C D} -> L →→hap (A · B) -> A →→st C -> B →→st D -> L →→st (C · D)
  st-abs : ∀{x A B} -> L →→hap (ƛ x A) -> A →→st B -> L →→st (ƛ x B)
  st-alpha : ∀{A' A} -> L →→st A' -> A' ∼α A -> L →→st A
\end{code}

\noindent The intention behind this relation is to characterize standard reduction sequences inductively. 
This definition allows us to perform as many \hap\ steps as we want. After that, if we reach a variable, then we are done since we cannot do any more reductions ({\tt st-var}). If the term is an application $A\ B$, then we can continue performing standard reductions on $A$ and then on $B$ ({\tt st-app}). Finally, if the term is an abstraction, we can continue performing standard reductions inside the body of the abstraction ({\tt st-abs}). Note that we are not forced to reduce all of the redexes that we encounter; given a redex, we can still apply  {\tt st-app} while skipping the head reduction.
The last constructor ({\tt st-alpha}) allows us to perform  $\alpha$-conversion. 
The preceding explanation may have shown that \emph{standard reductions} correspond to \emph{standard sequences} of reductions, i.e. that the former relation is included in the latter one. This is enough for proving the standardization theorem, as will be shown in the next subsection. We have further proven that actually the characterization of standard reduction sequences by the relation of standard reduction is complete, i.e. that the converse inclusion aldo holds.

The notion of standard reduction can be extended to substitutions. We say that  {\em substitution $\sigma $ standard-reduces  to $\sigma'\, $} ($\sigma \sst \sigma'$) \ if and only if for all variables $x$, $\sigma\ x \sta \sigma'\, x$.

\begin{code}
_→st_ : Σ → Σ → Set
σ →st σ' = (x : V) → σ x →→st σ' x
\end{code}


Reflexivity of \sta\ is proven by a direct induction on $M$ : Λ. 
\begin{code}
st-refl : ∀{M} -> M →→st M
\end{code}
Appending head reductions in applications at the beginning of a standard reduction results in a standard reduction. 
\begin{code}
hap-st→st : ∀{L M N} -> L →→hap M -> M →→st N -> L →→st N
\end{code}
\begin{proof}
By induction on the definition of $M \sta N$. 
\begin{itemize}
\item Case {\tt st-var}: We know that $M \hap x$. From this,  $L \hap M$ and the transitivity of \hap\  we conclude that $L \hap x$, and then  $L \sta x$ follows from {\tt st-var}. 
\item For the  {\tt st-app} case, assume $M \hap A\ B$, $A \sta C$ and $B \sta D$. From $L \hap M$ and $M \hap A\ B$ we conclude that $L \hap A\ B$ by  transitivity of \hap. Finally, from this plus $A \sta C$ and $B \sta D$, we conclude that $L \sta C\ D$ using  {\tt st-app}.
\item For the {\tt st-abs} case, we assume $M \hap\ ƛ x\ A$ and $A \sta B$. Similarly to the preceding case, we conclude that $L \hap\ ƛ x\ A$ from $L \hap M$, $M \hap\ ƛ x\ A$ and the transitivity of $\hap$. From this and $A \sta B$, we conclude $L \sta\ ƛ x\ B$ using {\tt st-abs}.\
\item For the {\tt st-alpha} case, we assume $M \sta A'$ and $A'\ \alfa A$. We use constructor {\tt st-alpha} applied to the induction hypothesis $L \sta A'$ and $A'\ \alfa A'$ to complete our goal.
\end{itemize}
\vspace*{-3.5ex}
\end{proof}

We can now use the preceding lemma to prove that substitution is preserved by the $\sta$ relation. This lemma is key to the proof and was originally stated by Kashima for single substitutions as: 
$M \sta N$ and $P \sta Q$\ $\Longrightarrow\ M[z := P] \sta N[z := Q]$.  
By taking the substitution to be an arbitrary (multiple)  $\sigma$ instead of the particular case where we replace just one variable $z$, we obtain a definition of substitution by structural recursion, and hence we can prove this result using just structural induction (see~\cite{COPELLO2016} for a detailed explanation). The substitution lemma is then stated as follows:

\begin{code}
st-substσ≅σ' : ∀{M N σ σ'} -> M →→st N -> σ →st σ' -> M ∙ σ →→st N ∙ σ'
\end{code}
\begin{proof}
By induction on the definition of $M \sta N$\
\begin{itemize}
\item Case {\tt st-var}: We have to prove $M\ \sust \sigma \sta x\ \sust \sigma'$ under the hypotheses $M \hap x$ and $\sigma\sst\sigma'$. \\
From {\tt hap-subst} applied to $M \hap x$ we know that $M\ \sust \sigma \hap x\ \sust \sigma$ and from the definition of $\sigma\sst\sigma'$ we get that $x\ \sust \sigma \sta x\ \sust \sigma'$. Therefore, from $M\ \sust \sigma \hap x\ \sust \sigma \sta x\ \sust \sigma'$ we conclude that $M\ \sust \sigma \sta x\ \sust \sigma'$ using {\tt hap-st→st}.\
\item Case {\tt st-app}: Assume $M \hap A\ B$,
$\sigma\sst\sigma'$,
$A\ \sust\ \sigma \sta C\ \sust\ \sigma' $
and $B\ \sust\ \sigma \sta D\ \sust\ \sigma' $.
We have to prove $M\ \sust\ \sigma \sta (C\ D)\ \sust\ \sigma'$. Now:\\
$M \hap A\ B$\\
$\Longrightarrow$ ({\tt hap-subst}) \\
$M\ \sust\ \sigma \hap (A\ B)\ \sust\ \sigma$\\
$\equiv$ (Def. $\sust$)\\
$M\ \sust\ \sigma \hap (A\ \sust\ \sigma)\ (B\ \sust\ \sigma)$\\
$\Longrightarrow$ ({\tt st-app} and hypothesis)\\
$M\ \sust\ \sigma \sta (C\ \sust\ \sigma')\ (D\ \sust\ \sigma')$\\
$\equiv$ (Def. \sust)\\
$M\ \sust\ \sigma \sta (C\ D)\ \sust\ \sigma'$.\
\item Case {\tt st-abs}: Assume $M \hap\ ƛ x\ A$, 
$A \sta B$ and $\sigma\sst\sigma'$.
We prove $M\ \sust\ \sigma \sta (ƛ x\ B)\ \sust\ \sigma'$:\\
$M \hap\ ƛ x\ A$\\
$\Longrightarrow$ ({\tt hap-subst}) \\
$M\ \sust\ \sigma \hap (ƛ x\ A)\ \sust\ \sigma$\\
$\equiv$ (Def $\sust$)\\
$M\ \sust\ \sigma \hap\ ƛ y_A\ (A\ \sust\ \sigma\aum(x,\ y_A))$ where  $y_A\ =\ \chi(\sigma,\ ƛ x\ A)$ (1).\\
Let $z = \chi (\iota \downharpoonright \ ((ƛ x\ A) \sust \sigma)\ ((ƛ x\ B)\ \sust\ \sigma'))$. Due to the definition of the choice function $\chi$, $z$ is fresh in every term and substitution involved. We can now prove that: \\
 $ƛ y_A\ (A\ \sust\ \sigma\aum(x,\ y_A))\ \alfa\ ƛz\ (A\ \sust\ \sigma\aum(x,\ z))$\\
$\Longrightarrow$ ({\tt hap-$\alpha$} and (1))\\
$M\ \sust\ \sigma \hap\ ƛ z\ (A\ \sust\ \sigma\aum(x,\ z))$ (2).\\
Note that by using multiple substitutions, our induction hypothesis is strong enough to allow us to use it with any pair of substitutions $\sigma, \sigma'$ as long as $\sigma \sst \sigma'$. Therefore, we can extract the following induction hypothesis from $A \sta B$:\\
$A\ \sust\ \sigma\aum(x,\ z) \sta B\ \sust\ \sigma'\aum(x,\ z)$ (3).\\
We can prove that $\sigma\aum(x,\ z) \sst \sigma'\aum(x,\ z)$  because the $\sta$ relation is reflexive (lemma {\tt st-refl}), so replacing $x$ for $z$ in both substitutions will preserve the $\sst$ relation.
So, from (2), (3) and constructor {\tt st-abs} we obtain that $M\ \sust \sigma \sta\  ƛz\ (B\ \sust \sigma'\aum(x,\ z))$ and  we obtain our thesis using {\tt st-alpha}, since\ $ƛz\ (B\ \sust \sigma'\aum(x,\ z))\ \alfa\ (ƛ x\ B)\ \sust \sigma'$, .\
\item Case {\tt st-alpha}: We assume that $M \sta N'$ and $N' \alfa N$ and want to prove that $M\ \sust\ \sigma \sta N\ \sust\ \sigma'$. From the induction hypothesis we get that $M\ \sust\ \sigma \sta N'\ \sust\ \sigma'$. In addition, we know that $N'\ \sust\ \sigma' \alfa N\ \sust\ \sigma'$, since they are equal ({\tt lemmaM∼M'→Mσ≡M'σ}) and \alfa\ is reflexive.  From these we obtain our goal using the {\tt st-alpha} rule.
\end{itemize}
\vspace*{-3.5ex}
\end{proof}
The following lemma states that if there is a standard reduction to a term that is a redex $(\lambda x.M)\ N$, then it is possible to construct a standard reduction to the contractum $M[x:=N]$, somehow ``inserting'' the contraction in a right place:
\begin{code}
st-abs-subst : ∀ {L M N x} -> L →→st (ƛ x M) · N -> L →→st (M [ x := N ])
\end{code}
\begin{proof}
From $L \sta (ƛ x\ M)\ N$ and the definition of \sta\ we know that $L \hap P\ N'$ for some $P$ and $N'$ such that $P \sta (ƛ x\ M)$ and $N' \sta N$. In addition, from $P \sta (ƛ x\ M)$ we know that $P \hap (ƛ x\ M')$ for some $M'$ such that $M' \sta M$. Then,\\
$P \hap (ƛ x\ M')$\\
$\Longrightarrow$ ({\tt hap-app-r})\\
$P\ N' \hap (ƛ x\ M')\ N'$ (1).\\
On the other hand,\\
\noindent $(ƛ x\ M')\ N'$\\
$\hap$ ({\tt α-star-singl} applied to constructor {\tt hap-head}) \\
$M'\ [x:=N']$\\
$\sta$ ({\tt st-substσ≅σ'} with $M'\sta M$ and $N' \sta N$)\\
$M\ [x:=N]$ (2).\\
From $L \hap P\ N'$, (1), (2) and the transitivity of $\hap$ we get that $L \hap M'\ [x:=N']$, and since $M'\ [x:=N'] \sta M\ [x:=N]$ we conclude that $L \sta M\ [x:=N]$ using lemma {\tt hap-st→st}.
\end{proof}
Using this result, we can now prove that any $\beta$-contraction can be also inserted into a standard reduction: 
\begin{code}
st-β→st : ∀{L M N} -> L →→st M -> M ⟶β N -> L →→st N
\end{code}
\begin{proof}
By induction on $M \sbet N$.
\begin{itemize}
\item The case {\tt outer-redex} follows directly from the previous lemma {\tt st-abs-subst}.
\item All the application cases are solved by simply using {\tt st-app} applied to the induction hypotheses. For example, if $M \sbet N$ was constructed using the rule {\tt appAbsL} then we know that $M=A \ C$, $N=B \ C$ and $(A\ C)\, \beta\, (B\ C)\, @\, (suc\ n)$, with $A\, \beta\, B\, @\, n$ for some $n$. Since $M$ is an application, $L \sta M$ must have been constructed using either the {\tt st-app} constructor or the {\tt st-alpha} constructor. We will deal with all the {\tt st-alpha} cases uniformly at the end, so let us focus on the {\tt st-app} case for now. We know that $L \hap A'\ C'$, $A'\sta A$ and $C' \sta C$. We want to prove that $L \sta B\ C$. From $A' \sta A$ and $A \sbet B$ we get that $A'\sta B$ by the induction hypothesis. Finally, we prove this case using the {\tt st-app} rule applied to $L \hap A'\ C'$, $A' \sta B$ and $C'\sta C$. The proofs for the other three application cases follow the same structure.
\item The {\tt abs} case also follows a similar pattern. We know that $\lambda x A \sbet \lambda x B$ where $A \sbet B$. Therefore, considering that $L \sta \lambda x A$ was constructed using the {\tt st-abs} rule, we have that $L \hap \lambda x A'$ and $A' \sta A$ for some $A'$. The induction hypothesis applied to $A'\sta A$ and $A \sbet B$ gives us  that $A' \sta B$, and we obtain our goal $L \sta \lambda x B$ using the {\tt st-abs} rule applied to $L \hap \lambda x A'$ and $A' \sbet B$.
\item In all the previous cases we ignored the case where {\tt L \sta M} \  was constructed using the {\tt st-alpha} constructor since we can prove this uniformly for all cases. We know that $L \sta M'$ and $M' \alfa M$. In order to use the induction hypothesis we would need to have that $M' \sbet K$ for some $K$. Since we know that $M \sbet N$ and $M' \alfa M$ we can use the $\alpha$-$\beta$ diamond property of Section~\ref{prelim} ({\tt lem-βα}), to obtain a term $K$ such that $M' \sbet K$ and $K \alfa N$, so we prove our goal using the {\tt st-alpha} rule.
\end{itemize}
\vspace*{-3.5ex}
\end{proof}

Finally, using this last result we can prove that if there is a sequence of $\beta$-reductions from $M$ to $N$, then there is also a standard reduction between those two terms. The proof is a direct induction on $M \bet N$:

\begin{code}
β→st : ∀{M N} -> M →→β N -> M →→st N
\end{code}

\subsection{Standard Sequences}

The next results show the relation between the reduction relations \leftmost, \hap\ and \sta\ with the existence of a standard reduction sequence.
Firstly notice that, since leftmost reductions always involve the reduction of redexes at position 0, then any sequence of leftmost reductions is a standard reduction sequence with lower bound 0. 
\begin{code}
nf→leftmost→seqβst : ∀{M N} -> M →→l N -> seqβ-st M N 0
\end{code}

As a corollary of this lemma and the fact that $M \hap N$ implies $M \leftmost N$ ({\tt lem-hap→l}), we obtain that
if $M \hap N$, then there is a standard reduction sequence from $M$ to $N$ with lower bound 0:
\begin{code}
hap→seqβst : ∀{M N} -> M →→hap N -> seqβ-st M N 0
\end{code}

The next result about {\sl seqβ-st} will be useful to prove the subsequent lemma. 
\begin{code}
abs-seq : ∀ {x M N n} -> seqβ-st M N n -> seqβ-st (ƛ x M) (ƛ x N) n
\end{code}

\begin{proof}
We proceed by induction on the definition of {\sl seqβ-st} $M\ N\ n$. 
\begin{itemize}
\item Case {\tt nil}: We know that {\sl seqβ-st} $M\ M\ 0$ and therefore we construct our goal, {\sl seqβ-st} $(\lambda x M)\ (\lambda x M)\ 0$, using {\tt nil}.  
\item Case {\tt $\alpha$-step}: We know that {\sl seqβ-st} $M\ K\ n$ for some $K$ such that $K \alfa N$. From the induction hypothesis we get that {\sl seqβ-st} $(\lambda x M)\ (\lambda x K)\ n$ and since $K \alfa N$ we can easily prove that $\lambda x K \alfa \lambda x N$. From this we prove the case using the {\tt st-alpha} constructor.
\item Case {\tt $\beta$-step}: We know that {\sl seqβ-st} $M\ K\ n$ for some $K$ such that $K\, \beta\, N\, @\, m$ with $n \leq m$. Similarly to the last case, the induction hypothesis tells us that {\sl seqβ-st} $(\lambda x M)\ (\lambda x K)\ n$ and from $K\, \beta\, N\, @\, m$ we can construct $(\lambda x K)\, \beta\, (\lambda x N)\, @\, m$ using rule {\tt abs}. Finally, we prove our goal using constructor {\tt $\beta$-step}.
\end{itemize}
\vspace*{-3.5ex}
\end{proof}

As for the \sta\ relation, 
if $M \sta N$ then there is a standard reduction sequence from $M$ to $N$, which we code in Agda as the existence of a lower bound for a standard reduction sequence: 

\begin{code}
st→seqβst : ∀{M N} -> M →→st N -> Σ ℕ (\n -> seqβ-st M N n)
\end{code}

\begin{proof}
By induction on the definition of $M \sta N$. 
\begin{itemize}
\item The case {\tt st-var} can be easily proven using lemma {\tt hap→seqβst}: since $M \hap x$, then there is a standard reduction sequence (with lower bound 0) from $M$ to $x$. 
\item Similarly, the case {\tt st-abs} also relies in this lemma, and the induction hypothesis: we know from {\tt hap→seqβst} that there is a standard reduction sequence with lower bound 0 from $M$ to $\lambda x. A$; the induction hypothesis tells us that there exists a natural number $n$ such that there is a reduction sequence from $A$ to $B$ with lower bound $n$. Therefore, using lemma {\tt abs-seq} we conclude that there must be a standard reduction sequence from $M$ to $B$ with lower bound $n$ since $0 \leq n$. 
\item The case for {\tt st-app} is slightly trickier since the lower bound that exists depends on certain characteristics of the terms involved:
If $M \sta N$ was constructed using the constructor {\tt st-app}, that means that for some terms $A$, $B$ , $C$ and $D$:
(1) $M \hap (A\ B)$, (2) $A \sta C$ and (3) $B \sta D$. We need to prove that there is a standard reduction sequence from $M$ to $(C\ D)$. Using the induction hypotheses, let $m$ and $n$ be the lower bounds for the standard reduction sequences from $A$ to $C$ and from $B$ to $D$ respectively:
\begin{enumerate}
\item If $C$ is not an abstraction, and $B \alfa D$\footnote{This is a possible scenario, since \sta\ includes \alfa.}, then the lower bound for the standard reduction sequence will be $m$.
\item If $C$ is not an abstraction, and $B \not\alfa  D$, then the lower bound for the standard reduction sequence will be $n + {\tt countRedexes}\ C$.
\item If $C$ is an abstraction, and $B \not\alfa  D$, then the lower bound for the standard reduction sequence will be $suc\,(n + {\tt countRedexes}\ C\,)$.
\item If $C$ is an abstraction, and $B \alfa D$, then  we need to do some further case analysis using the following lemma:
\begin{code}
lem-seq-appACBC-abs : ∀ {A C B n} -> seqβ-st A C n -> isAbs C
  -> (seqβ-st (A · B) (C · B) n) ∨ (seqβ-st (A · B) (C · B) (suc n))
\end{code}
Note that if a reduction sequence ends in an abstraction, by appending the same application (or an $\alpha$-equivalent one) to all of the terms in the sequence, the lower bound will remain the same if and only the abstraction is generated in the last beta step of the sequence and therefore does not affect the redex count in $\beta$-reductions. However, if the abstraction appears before that, then the lower bound of the reduction sequence must be increased by one, since a new redex at position 0 is formed. From this lemma we conclude that the lower bound must be either $m$ or $suc\ m$ for this case. 
\end{enumerate}
\item Finally, for the {\tt st-alpha} case, we have that $M \sta N'$ and $N' \alfa N$ and we want to prove the existence of a standard reduction sequence from $M$ to $N$. The induction hypothesis gives us a standard reduction sequence from $M$ to $N'$ and we can directly perform an alpha step to $N$ by using the {\tt $\alpha$-step} constructor from {\tt seq$\beta$-st}.
\end{itemize}
\vspace*{-3.5ex}
\end{proof}

The Standardization Theorem finally follows from this last result and lemma {\tt β→st} that states that $M \bet N$ implies $M \sta N$:

\begin{code}
standardization : ∀{M N} -> M →→β N -> Σ ℕ (λ n -> seqβ-st M N n)
standardization M→→βN = st→seqβst (β→st M→→βN)
\end{code}



\section{The Leftmost Reduction Theorem}
\label{leftmost}

A quite relevant corollary of the Standardization Theorem is the Leftmost Reduction Theorem, which states that if a term $M$ has a normal form, then the leftmost-outermost reduction strategy will find it. 
In the present section we show how this property can be derived from standardization. It is worth noticing that this proof was developed as part of the present work and is not present in  Kashima's article. 

We can directly characterize a term in normal form as one without redexes using the {\tt countRedexes} function from section~\ref{prelim}:

\begin{code}
nf : Λ -> Set
nf M = countRedexes M ≡ 0
\end{code}
and now we can state the aforementioned property as the following lemma:

\begin{code}
leftmost-nf : ∀{M N} -> M →→β N -> nf N -> M →→l N
\end{code}

In order to prove this result, we must first consider some lemmas. 
The first one states that the number of redexes of two $\alpha$-equivalent terms is the same, which easily follows by induction on $M \alfa N$:
\begin{code}
α→sameRedexCount : ∀ {M N} -> M ∼α N -> countRedexes M ≡ countRedexes N
\end{code}

The second lemma states that if a term $M$ $\beta$-reduces to a term $N$ in normal form, then the contracted redex must be the leftmost redex of $M$, i.e., the one at position zero:
\begin{code}
nf→l : ∀{M N n} -> M β N @ n -> nf N -> n ≡ 0
\end{code}
\begin{proof}
We proceed by induction on $M\, \beta\, N\, @\, n$ 
\begin{itemize}
\item Case {\tt outer-redex}: we have that $(\lambda x A)\, B\, \beta\, B[x:=A]\, @\, 0$. Our goal follows directly since rule {\tt outer-redex} contracts the redex at position $0$.
\item Case {\tt appNoAbsL}: we have that  $(A\, C)\, \beta\, (B\, C)\, @\, n$ where $A\, \beta\, B\,@\, n$, $A$ is not an abstraction and $(B\ C)$ is in normal form. From this, we know that ${\tt countRedexes}\ B + {\tt countRedexes}\ C \equiv 0$, and therefore {\tt countRedexes $B$ $\equiv$ 0} which allows us to use the induction hypothesis with $A\, \beta\, B\, @\,n$ and conclude that $n \equiv 0$.
\item Case {\tt appAbsL}: we have that $(A\, C)\, \beta\, (B\, C)\, @\, (suc\, n)$ where $A\, \beta\, B\, @\, n$, $A$ is an abstraction and $(B\ C)$ is in normal form. Since $(B\ C)$ is in normal form $B$ cannot be an abstraction, but this is a contradiction since $A\, \beta\, B\,@\, n$ and $A$ is an abstraction because contracting a redex in an abstraction always results in an abstraction (rule {\tt abs}).
\item Case {\tt appNoAbsR}: we have that  $(C\, A)\, \beta\, (C\, B)\, @\, ({\tt countRedexes}\, C + n)$  where  $A\, \beta\, B\, @\, n$, $C$ is not an abstraction and $(C\ B)$ is in normal form. From this we know that ${\tt countRedexes}\ C \equiv 0$ and ${\tt countRedexes}\ B \equiv 0$. From the induction hypothesis we have that $n \equiv 0$, and since ${\tt countRedexes}\ C \equiv 0$, $n+ {\tt countRedexes}\ C \equiv 0$.
\item Case {\tt appAbsR}: we have that  $(C\, A)\, \beta\, (C\, B)\, @\, suc\,({\tt countRedexes}\, C + n)$  where  $A\, \beta\, B\, @\, n$, $C$ is an abstraction and $(C\ B)$ is in normal form. However, this is a contradiction since $(C\ B)$ cannot be in normal form if $C$ is an abstraction.
\item Case {\tt abs}: we have that $\lambda x A\ \beta\ \lambda x B\ @\ n$ where $A\ \beta\ B\ @\ n$ and $\lambda x B$ is in normal form. From this we know that $B$ must be in normal form too and we can call the induction hypothesis for $A\ \beta\ B\ @\ n$, concluding that $n\equiv 0$.
\end{itemize}
\vspace*{-3.5ex}
\end{proof}

Finally, notice that a standard sequence with lower bound $0$ must be a sequence of leftmost reductions, since all of the $\beta$-steps must involve the contraction of the redex at position 0, i.e. a leftmost reduction.

\begin{code}
seqβ0→l : ∀ {A B} -> seqβ-st A B 0 -> A →→l B
\end{code}
The proof follows by a direct induction on {\sl seqβ-st} $A\ B\ 0$.\\

Lets now turn our attention to the  main lemma:

\begin{code}
seqst→l : ∀{M N n} -> seqβ-st M N n -> nf N -> M →→l N
\end{code}

\begin{proof}
We proceed by induction on the definition of {\sl seqβ-st} $M \ N\ n$.
\begin{itemize}
\item Case {\tt nil}: We have that {\sl seqβ-st} $A\ A\ 0$ \  and we need to prove that $A \leftmost A$, which follows by constructor {\tt refl}.
\item Case {\tt $\alpha$-step}: We have that {\sl seqβ-st} $A\ B'\ n$ with $B' \alfa B$ and \ {\tt nf} $B$. From this, we have that {\tt countRedexes} $B' \equiv 0$ by lemma {\tt α→sameRedexCount}  and therefore, using the induction hypothesis we obtain {\tt A $\leftmost$ B'}. Finally, we construct our goal using rule {\tt $\alpha$-step}.
\item Case {\tt $\beta$-step}: We have {\sl seqβ-st} $A\ B'\ n$ and $B'\,\beta\,B\,@\,n'$, where $n \leq n'$ and {\tt countRedexes} $B \equiv 0$. Using lemma {\tt nf→l} we have that $n' \equiv 0$, and so $n \equiv 0$. We then apply lemma {\tt seqβ0→l} to {\sl seqβ-st} $A\ B'\ 0$ and get that $A \leftmost B'$. Note that since $n' \equiv 0$, we also have that $B' \sleftmost B$. Finally, from $A \leftmost B'$ and $B' \sleftmost B$ we conclude that $A \leftmost B$ using rule {\tt append}. 
\end{itemize}
\vspace*{-3.5ex}
\end{proof}

Finally, if we have that $M \bet N$, the Standardization Theorem lets  us  conclude  that  there  exists  a standard reduction sequence from $M$ to $N$. Therefore, the desired property follows directly combining this result and  the previous lemma:

\begin{code}
leftmost-nf : ∀{M N} -> M →→β N -> nf N -> M →→l N
leftmost-nf M→→βN crN≡0 = seqst→l (proj2 (standardization M→→βN)) crN≡0
\end{code}

\section{Conclusions}
\label{concl}

In this work we have extended some metatheoretical results from~\cite{COPELLO2016} by formalizing a proof of the Standardization Theorem in Lambda Calculus using Constructive Type Theory. We use a concrete approach to $\lambda$-terms and the notion of multiple substitution. The latter enables us to proceed by structural induction only, producing proofs that are easy to follow, yet fully formal. 
This work has also served to showcase the usefulness of the library produced in~\cite{ernius:lib} and its suitability for the formalization of other metatheoretical properties of the Lambda Calculus. It is worth highlighting that the definitions and lemmas used to handle  syntax and substitutions did not need to be modified or extended in any way
and could be rapidly put into use by a programmer with a minimal training in Agda, namely the first author while working on his Master's thesis~\cite{copes:thesis}. The Agda code reported in this paper is 890 lines long.


Other efforts to formalize Kashima's proof in the literature include one by Guidi in Matita~\cite{guidi:standard} and another one by  Vyšniauskas and  Emerich in Coq~\cite{standard:coq}. However, what sets our development apart from these efforts is the use of a concrete syntax with names and our definition of multiple substitution. While this allows us to prove our lemmas using a clean structural inductive argument, they use a nameless syntax based on de Brujin indexes which results in some inductions being done on the size of the $\lambda$-terms. Another effort worth highlighting is that of McKinna and Pollack, who formalized a proof of the Standardization Theorem due to Takahashi~\cite{takahashi} using the LEGO proof assistant~\cite{McKinna1999}. 

Within the chosen syntax approach, we have to consider the work by Vestergaard and Brotherston~\cite{vestergaard} which uses modified rules of $\alpha$-conversion and $\beta$-reduction based on unary substitution to formally prove the Church-Rosser theorem in Isabelle-HOL.
Substitution does not proceed in cases of capture and they use explicit $\alpha$-conversion to perform the renaming achieved by our substitution.
As a consequence, their development requires an administrative layer of reasoning for showing that $\alpha$-conversion and $\beta$-reduction interact correctly. This consists in a rather complex definition of a new auxiliary relation for $\alpha$-conversion, which we do not need. 

In addition to proving the Standardization Theorem, Kashima proves a few other interesting results which could be a good follow up to the present work, e.g. the {\em Quasi-Leftmost Reduction Theorem}. An infinite $\beta$-reduction sequence is called quasi-leftmost if it contains infinitely many leftmost reduction steps $\sleftmost$. As a corollary of the Standardization Theorem it can be proved that if $M$ has a $\beta$-normal form, then there is no infinite quasi-leftmost $\beta$-reduction sequence from $M$. 




\bibliographystyle{eptcs}
\bibliography{bibliography}

\end{document}